\documentclass[reprint,superscriptaddress,floatfix,longbibliography,tightenlines,nobibnotes,nofootinbib,amsmath,amssymb,aps,pra]{revtex4-1}
\usepackage{graphicx}
\usepackage{color}
\usepackage{bm}
\usepackage{dcolumn}
\usepackage{textcomp}
\usepackage{hyperref}
\hypersetup{colorlinks=true,urlcolor= blue,citecolor=blue,linkcolor= blue,bookmarks=true,bookmarksopen=false}
\usepackage{verbatim}
\usepackage{blindtext}
\usepackage{natbib}
\setcitestyle{numbers,square}
\usepackage{makecell}
\usepackage{amsmath}
\usepackage{amsfonts}
\usepackage{amssymb}
\usepackage{comment}
\usepackage{mathtools}
\usepackage{braket}
\usepackage{soul}
\begin{document}

\title{
Berry curvature, orbital magnetization, and 
Nernst effect in biased bilayer WSe$_2$}
\author{Vassilios Vargiamidis}
\email{V.Vargiamidis@warwick.ac.uk}
\affiliation{School of Engineering, University of Warwick, Coventry, CV4 7AL, United Kingdom}
\author{P. Vasilopoulos }
\email{p.vasilopoulos@concordia.ca}
\affiliation{Department of Physics, Concordia University, 7141 Sherbrooke Ouest, Montreal, Quebec H4B 1R6, Canada}
\author{M. Tahir}
\email{tahir@colostate.edu}
\affiliation{Department of Physics, Colorado State University, Fort Collins, CO 80523, USA}
\author{Neophytos Neophytou}
\email{N.Neophytou@warwick.ac.uk}
\affiliation{School of Engineering, University of Warwick, Coventry, CV4 7AL, United Kingdom}

\begin{abstract}

A valley-contrasting Berry curvature in bilayer transition metal dichalcogenides  with spin-orbit coupling can generate valley magnetization when 
the inversion symmetry is broken, for example, by an 
electric field, regardless of time-reversal symmetry. A nontrivial Berry curvature can also lead to 
anomalous transport responses, such as the anomalous Hall effect  and the anomalous Nernst effect. 
Applied to a bilayer WSe$_2$, an 
electric field can tune the Berry curvature and orbital magnetic moment, which has important consequences for the orbital magnetization 
and  the anomalous Nernst responses. The 
orbital magnetization 
and its two contributions, one due to the magnetic moment and one due to the Berry curvature, are calculated and interpreted in terms of opposite circulating currents of the bands in the two 
 layers. The valley anomalous Nernst coefficient  and spin Nernst coefficient are also calculated. We find that a finite electric field leads to peaks and dips in the Nernst responses that have the  signs 
of the Berry curvatures of the bands and are proportional to their magnitudes; it also enhances the valley Nernst responses. These experimentally verifiable findings may be promising for caloritronic applications.
	
\end{abstract}

\maketitle
\date{\today}

\section{\textbf{Introduction}}

In two-dimensional (2D) materials and transition metal dichalcogenides (TMDCs) with hexagonal structure, an electron not only has spin but also a valley degree of freedom, which acts as a pseudospin. Phenomena such as valley polarization and valley- and spin-Hall effects have been discussed for the $K$ and $K^{\prime} = -K$ Dirac valleys at opposite corners of the Brillouin zone \cite{xiao12,cao12,mak14,shan13,lee16,habe17}. The realization of these effects is based on the control 
of properties that differ between the two valleys \cite{xiao07}, in particular the magnetic moment $(\mathbf{m})$ and the Berry curvature $(\mathbf{\Omega})$. For spinful electrons time-reversal ($TR$) symmetry dictates that $\mathbf{\Omega}$ has the same magnitude but opposite sign in the two valleys for opposite spin states, i.e., $\mathbf{\Omega}_{n, \sigma}(K) = -\mathbf{\Omega}_{n, -\sigma}(-K)$, where $n$ is a band index and $\sigma$ labels a spin state. Space-inversion ($P$) symmetry requires them to have the same sign, i.e., $\mathbf{\Omega}_{n, \sigma}(K) = \mathbf{\Omega}_{n, \sigma}(-K)$. If both $TR$ and $P$ are good symmetries, $\mathbf{\Omega}_{n, \sigma}(K) = - \mathbf{\Omega}_{n, -\sigma}(K)$. The symmetry properties for $\mathbf{m}$ are the same as for $\mathbf{\Omega}$. Therefore, in order to have nonzero $\mathbf{\Omega}$ and $\mathbf{m}$ one of the two symmetries must be broken. A necessary condition for valley-contrasting $\mathbf{m}$ and $\mathbf{\Omega}$ is the $P$ symmetry breaking \cite{xiao07}, independent of the   
$TR$ symmetry. The Berry curvature is also allowed to be nonzero in the presence of spin-orbit coupling (SOC). In this case the spin-dependent spatial states with opposite spins have different orbital wave functions and there is no requirement that $\mathbf{\Omega}_{n, \sigma}(K)$ be zero.

The Berry curvature can be described as a pseudo-magnetic field in  reciprocal space which drives the carriers in 
 its direction. A direct consequence is the valley-Hall effect (VHE) \cite{xiao12,xiao07}, in which electrons at the two valleys drift to opposite edges of the material in the presence of an in-plane electric field due to the equal but opposite Berry curvatures at the two valleys \cite{xu14}. A nonzero Berry curvature in TMDCs also leads to various anomalous transport phenomena, such as the anomalous Hall effect (AHE) \cite{habe17} and the anomalous Nernst effect 
 (ANE)  \cite{xiao06}. TMDCs also exhibit strong SOC which, together with the intrinsic broken $P$ symmetry, leads to a coupling of the spin and valley degrees of freedom \cite{xiao12,gong13}. On the other hand, the layer degree of freedom 
 in bilayer systems  can be described as a pseudospin. Pseudospin up (down) refers to the state where the charge carrier is located in the upper (lower) layer. For a SOC strength larger than the interlayer hopping, a carrier is localized in either the upper or lower layer resulting in spin-layer locking effect \cite{jones14} in a particular valley, and therefore the spin remains a good quantum number.

Bilayer TMDCs are AB stacked, i.e., one monolayer sits on another but rotated by 180$^{\circ}$. 
In bilayer WSe$_2$ the Se atoms in the upper layer sit on top of the W atoms of the bottom one. Pristine bilayers are therefore inversion symmetric. However, the $P$ symmetry can be broken by applying an electric field perpendicular to a bilayer; this  causes a potential difference between the two layers and leads to valley-contrasting Berry curvature and magnetic moment. In fact, the emergence of valley-contrasting physical properties is a generic consequence of the $P$ symmetry breaking in 2D hexagonal lattices. In such systems, the effects of the Berry curvature give rise to topological, electric \cite{xiao07}, and thermoelectric \cite{xiao06} transport phenomena. Important developments on thermoelectric transport were based on the Berry-phase correction to the  orbital magnetization (OM)
at finite temperatures \cite{xiao06}, and its role on the intrinsic Hall current. In fact, the Berry-phase correction term, which is of topological nature, eventually enters the transport current and leads to an anomalous Nernst conductivity. Topological thermoelectric transport has been studied previously in single and bilayer graphene \cite{sarma09}, while for monolayer TMDCs a mechanism for generation of pure spin current via the spin-Nernst effect was described in Ref.~\cite{jauho15}. Recently, the Nernst response of monolayer and bilayer TMDCs was investigated in the presence of the Rashba SOC \cite{sharma18}. The ANE was also investigated experimentally \cite{liang17} and theoretically \cite{sharma17} in Dirac and Weyl semimetals.

The purpose of this work 
is to investigate the Berry curvature, the OM, 
and the topological Nernst effect in bilayer WSe$_2$ with broken $P$ symmetry via gating. In the presence of SOC the application of an electric field causes a splitting of the spin-degenerate bands that  now 
have opposite spin polarizations 
at the two valleys as enforced by $TR$ symmetry. After  briefly presenting 
the band structure, we calculate the Berry curvature and orbital magnetic moment and explore how they are affected by the electric field.  
We find that, for either 
valley the electric field tunes the Berry curvature of the spin-down bands but  barely affects that of the spin-up ones. The same holds 
for the orbital magnetic moment. 
We then investigate the OM 
for each valley. 
When the  electric field is  absent
 the OM 
for each 
valley vanishes as a consequence of the simultaneous presence of $TR$ and $P$ symmetries. When it's present though, 
the OM 
varies linearly with the chemical potential $E_F$ but is constant in the band gap. The 
electric field also leads to nonzero Nernst responses due to the nontrivial Berry curvature of the bands. We evaluate the valley anomalous Nernst coefficient (ANC) and spin-Nernst coefficient (SNC), as functions of $E_F$, 
for zero and finite electric fields at various temperatures. 
The electric field leads to peaks and dips in the Nernst responses, that are proportional to the magnitudes 
of the Berry curvatures, 
and to enhanced  valley Nernst signals.

 In Sec.~II we present the Hamiltonian of a bilayer WSe$_2$, briefly discuss its band structure, and evaluate 
 the 
 matrix elements of the velocity operator. In Sec.~III we discuss the effects of the electric field on the Berry curvature and orbital magnetic moment, and make a comparison with two decoupled monolayers. In Sec.~IV we discuss the OM 
 and in Sec.~V we explore the ANE. 
 We summarize and conclude in Sec.~VI.

\section{Theoretical model}

\textit{Ab initio} studies of the band structure revealed that edges of valence and conduction bands in bilayer TMDCs near the $K$ points are dominantly comprised of $d_{xy}$, $d_{x^2 - y^2}$, and $d_{z^2}$ states. As a result, the effective Hamiltonian can be constructed by adding interlayer coupling \cite{gong13, wu13, jones14, fang15} to the $\mathbf{k} \cdot \mathbf{p}$ model of monolayers established in Ref.~\cite{xiao12}. We consider a bilayer WSe$_2$ with an interlayer coupling $\gamma$ and an applied 
electric field which induces a 
potential energy difference $2V$ between the 
 layers. The Hamiltonian near the $K$ and $K^{\prime}$ valleys reads
\begin{equation}
H = \left(
\begin{array}
[c]{cccc}%
\delta_1
 &  \hbar v_F k_-  &  0  &  0\\
\hbar v_F k_+  &  \delta_2
 &  0   &  \gamma\\
0  &  0  &  -\delta_2+\delta_\lambda 
&  \hbar v_F k_+\\
0  &  \gamma  &  \hbar v_F k_-  &  -\delta_1-\delta_\lambda
\end{array}
\right), \label{eq1}%
\end{equation}
where 
\begin{equation}
\delta_1 = \Delta + \tau s_z \lambda_c +V, \,
\delta_2 = - \Delta + \tau s_z \lambda_v +V, \delta_\lambda=\tau s_z (\lambda_v-\lambda_c), 
\label{eq9}
\end{equation}
Further, $k_{\pm} = \tau k_x \pm i k_y$, where $\mathbf{k}$ is the relative wave vector with respect to the $K$ points, and $\tau = \pm 1$ for the $K$, $K^{\prime}$ valleys of the bilayer bands. Also, $v_F = 5 \times 10^5$m/s is the Fermi velocity, $2 \Delta = 1.7$eV is the band gap of monolayer WSe$_2$ \cite{gong13}, and $\gamma = 0.067$eV is the interlayer hopping for holes. The interlayer hopping for electrons vanishes at $K$ points due to the symmetry of the $d_{z^2}$ orbital. Also, $s_z$ denotes the Pauli matrix for the $z$ component of the spin. From Eq.~(\ref{eq1}) we see that $[s_z, H] = 0$, and thus $s_z$ is a good quantum number. The spin-up ($\uparrow$) and spin-down ($\downarrow$) states corresponding to $s_z = \pm 1$ are decoupled in bilayers, as interlayer coupling conserves spin. Further, $\lambda_c = 7.5$meV is the SOC for electrons and $\lambda_v = 112.5$meV that for holes. Note that the diagonal elements in Eq.~(\ref{eq1}) involve the terms $\tau s_z \lambda_{c}$ and $\tau s_z \lambda_{v}$,  due to the fact that the heavy W atoms induce a strong SOC with large spin-splitting (especially for holes) leading to spin-valley coupling. This necessarily leads to a simultaneous spin flip in addition to large momentum transfer ($K \longleftrightarrow K^{\prime}$) during scattering of charge carriers between valleys, which is of great importance for the realization of the valley spin valve in 2D materials \cite{tao19}. The SOC in Eq.~(\ref{eq1}) is due to terms $H_{so, c} = \lambda_c \tau s_z \mu_z$ and $H_{so, v} = \lambda_v \tau s_z \mu_z$ in the conduction and valence band of the bilayer, respectively, in the single layer Hamiltonian of Ref.~\cite{xiao12}. Here, the Pauli matrix $\mu_z$ is the layer pseudospin which indicates that the SOC has a different sign \cite{gong13} in the two layers. This originates from the 180$^{\circ}$ rotation of the two layers with respect to each other. The electrostatic potential $V$ in Eq.~(\ref{eq1}) is due to the Hamiltonian $H_V = V \mathbf{1}_{s_z} \mathbf{1}_{\sigma} \mu_z$ where $\mathbf{1}_{s_z}(\mathbf{1}_\sigma)$ denotes the identity matrix in the $s_z ( \sigma )$ space with $\sigma_i (i=x,y)$ the Pauli matrices for the two basis functions of the energy bands of a monolayer \cite{xiao12}.

\subsection{Eigenvalues and eigenfunctions}

To obtain the eigenvalues  and eigenstates of Eq.~(\ref{eq1}) we first write $k_{\pm} = \tau \vert \mathbf{k} \vert \text{e}^{\pm i \tau \varphi_\mathbf{k}}$ where $\tan(\tau \varphi_{\mathbf{k}}) = \tau k_y / k_x$ and $\vert \mathbf{k} \vert = k = \left( k_x^2 + k_y^2 \right)^{1/2}$. Then setting equal to zero the determinant corresponding to the  Hamiltonian (\ref{eq1})
leads to the 
quartic equation for the eigenvalues $E$
\begin{equation}
E^4 +a_1 E^2 + a_2 E + a_3 = 0 ;
\label{eq3}%
\end{equation}
with $ \epsilon_k = \hbar v_F k$ the coefficients $a_1, a_2, a_3$ are given by
\begin{eqnarray}
\hspace*{-1.4cm}
a_1 &=& - \delta_1^2- \delta_2^2
-(\delta_1 - \delta_2)\delta_\lambda-\delta_\lambda^2 -2\epsilon_k^2
- \gamma^2,\\* 
%
\hspace*{-1.4cm}
a_2 &=&( \delta_1 + \delta_2)[( \delta_1 - \delta_2)\delta_\lambda+\delta_\lambda^2]
+\gamma^2 ( \delta_1 - \delta_2 +\delta_\lambda),\\*
\notag
\hspace*{-1.4cm}
a_3 &=& ( \delta_1 \delta_2 - \epsilon_k^2 ) ( \delta_1 + \delta_\lambda ) ( \delta_2 - \delta_\lambda ) - \epsilon_k^2 \delta_1 \delta_2 \\*
&&+ \gamma^2 \delta_1 (\delta_2 -\delta_\lambda)+ \epsilon_k^4  ,
\label{eq6}%
\end{eqnarray}
%
%

The solutions of Eq.~(\ref{eq3}) are
\begin{eqnarray}
\notag
E_{\lambda \mu}^{\tau s_z} ( k ) &=& \frac{1}{2 \sqrt{3}} \Big\{ \lambda \Big[ - 2 a_1 + B + \frac{A}{2^{1/3}} \Big]^{1/2} + \mu \Big[ - 4 a_1  \\*
&- B& -\frac{A}{2^{1/3}} 
- \lambda \frac{6 \sqrt{3} a_2}{\sqrt{- 2 a_1 + B + A/2^{1/3}}} \Big]^{1/2} \Big\} ,
\label{eq11}%
\end{eqnarray}
with $\lambda = + 1 (-1)$  for the conduction (valence) band. The pseudospin $\mu$  is the layer degree of freedom: $\mu = + 1 (-1)$ for the upper (lower) layer. The constants $A,\,B$ are given in terms of the constant $C=2 a_1^3 + 27 a_2^2 - 72 a_1 a_3  $\,\, as
%
%
%
%
\begin{eqnarray}
A &=& \Big[ C +
[C^2 - 4 ( a_1^2 + 12 a_3 )^3]^{1/2} \Big]^{1/3}\\*
B &=& 2^{1/3} ( a_1^2 + 12 a_3)/A.
\label{eq14}%
\end{eqnarray}
\begin{figure}[ptb]
\vspace*{-0cm}
\begin{center}
\includegraphics[height=4.4cm, width=8.6cm ]{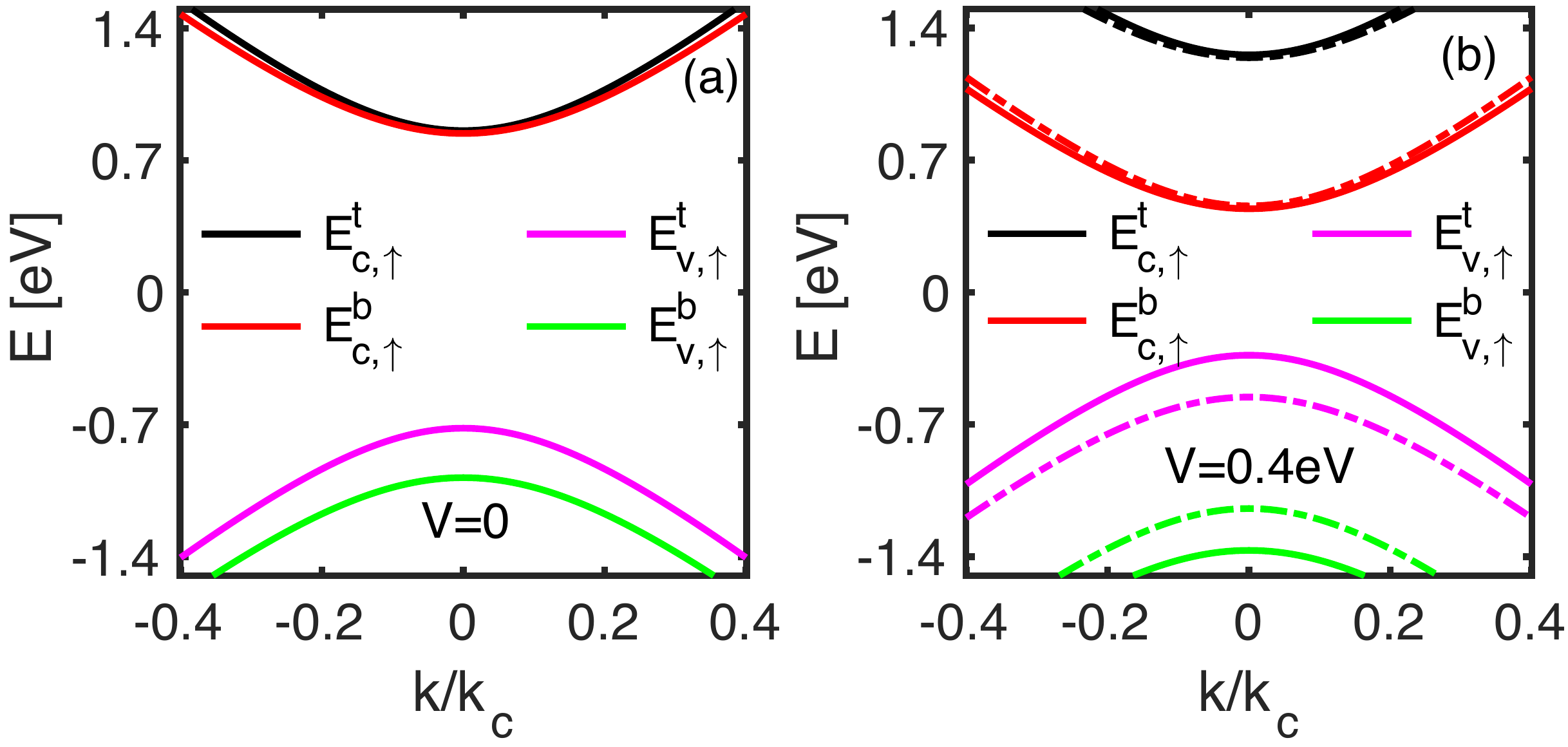}
\end{center}
\vspace*{-0.55cm} \caption{(Colour online) Energy dispersion near the $K$ valley for bilayer WSe$_2$ in the absence (left panel) and presence (right panel) of a perpendicular electric field corresponding to a 
potential energy $V=0.4$eV. The solid curves are for spin-up bands and the dash-dotted ones  for spin-down bands which are not labelled for clarity. Notice the enhanced splitting of the valence spin subbands.} 
\label{fig:fig1}%
\end{figure}

In Fig.~1 we show the energy dispersion $E_{\lambda \mu}^{\tau s_z} (k)$ of bilayer WSe$_2$ for the $K$ valley versus $k/k_c$ where $k_c = \pi / a$ and $a = 3.32 \text{\AA}$ is the lattice constant of monolayer WSe$_2$. The spin-up, conduction (valence) band energies in the top (bottom) layer are denoted by $E_{c, \uparrow}^{t} (E_{c, \uparrow}^{b})$ and $E_{v, \uparrow}^{t} (E_{v, \uparrow}^{b})$. In the absence of an electric field [see Fig.~1(a)] Kramer's degeneracy is established [$E_{n, \uparrow}^{\ell} ( \mathbf{k} ) = E_{n, \downarrow}^{\ell} ( \mathbf{k} )$, where $n = c ( v )$ and $\ell = t ( b )$] by the combination of $TR$ [$E_{n, \uparrow}^{\ell} ( \mathbf{k} ) = E_{n, \downarrow}^{\ell} ( -\mathbf{k} )$] and $P$ [$E_{n, \uparrow}^{\ell} ( \mathbf{k} ) = E_{n, \uparrow}^{\ell} ( -\mathbf{k} )$] symmetries. In the conduction band the splitting between the two-fold degenerate levels is essentially given by the SOC strength $2 \lambda_c$ of monolayer TMDCs at $\mathbf{k} = 0$. In the valence band the energy splitting would be $2 \gamma$ if the SOC was absent. However, the presence of the strong SOC for holes causes significant increase in the energy splitting from $2 \gamma$ to $2 \sqrt{\gamma^2 + \lambda_v^2}$. For finite 
$V$ 
the spin degeneracy is lifted, due to the breaking of the 
$P$ symmetry, and leads to finite splitting of the spin subbands, as shown in Fig.~1(b). The spin-up bands are shown by the solid lines while the spin-down ones by the dash-dotted curves; the latter are not labelled for clarity. The spin-splitting due to a finite $V$ is more pronounced for the valence bands 
than for the conduction bands. This is because 
the electric field enhances the effect of SOC resulting in larger spin and layer splittings. Notice also that the band gap is reduced as $V$ increases. For the $K^{\prime}$ valley (not shown) the spin subbands are reversed.

The eigenstates of $H$ are four-component spinors,
$\Psi =\left(
\begin{array}
[c]{c}%
\phi_{1}\,,
\phi_{2}\,,
\chi_{1}\,,
\chi_{2}%
\end{array}
\right)^T$ 
%
where $T$ denotes the transpose. Following a standard diagonalization procedure we obtain

\begin{equation}
\Psi_{\lambda \mu \mathbf{k}} ( \mathbf{r} ) = N_{\lambda \mu \mathbf{k}} \left(
\begin{array}
[c]{c}%
\vspace{0.07in} \frac{\epsilon_k \tau \text{e}^{-i \tau \varphi_{\mathbf{k}}}}{E_{\lambda \mu \mathbf{k}} - \delta_1} \eta_{\lambda \mu \mathbf{k}}\\
\vspace{0.07in} \eta_{\lambda \mu \mathbf{k}}\\
\vspace{0.07in}  \frac{\epsilon_k \tau \text{e}^{i \tau \varphi_{\mathbf{k}}}}{E_{\lambda \mu \mathbf{k}} + \delta_2-\delta_\lambda} \\
1%
\end{array}
\right) \frac{\text{e}^{i \mathbf{k} \cdot \mathbf{r}}}{\sqrt{S}}  ,\label{eq18}%
\end{equation}
where we suppressed the superscripts $\tau$ and $s_z$ for clarity;  $E_{\lambda \mu \mathbf{k}}$ are the energy eigenvalues given in Eq.~(\ref{eq11}). We have also defined the auxiliary quantity $\eta_{\lambda \mu \mathbf{k}}$ as
\begin{equation}
\eta_{\lambda \mu \mathbf{k}} = \frac{\left( E_{\lambda \mu \mathbf{k}}+\delta_2 - \delta_\lambda \right) \left( E_{\lambda \mu \mathbf{k}}+\delta_1 + \delta_\lambda\right) - \epsilon_k^2}{\gamma \left( E_{\lambda \mu \mathbf{k}}+\delta_2 -\delta_\lambda\right)}  .
\label{eq19}%
\end{equation}
Note that $\eta_{\lambda \mu \mathbf{k}}$ is purely real. Also, with $S$  the area of the sample,  the normalization factor $N_{\lambda \mu \mathbf{k}}$ 
is found to be
\begin{eqnarray}
\nonumber \hspace*{0in} N_{\lambda \mu \mathbf{k}} = \Big\{ \eta_{\lambda \mu \mathbf{k}}^2 \Big[ 1 + \frac{\epsilon_k^2}{( E_{\lambda \mu \mathbf{k}} - \delta_1 )^2} \Big] 
\\* &&\hspace*{-1.3in} + 1+\frac{\epsilon_k^2}{( E_{\lambda \mu \mathbf{k}} + \delta_2-\delta_\lambda )^2} \Big\}^{-1/2} ,
\label{eq20}%
\end{eqnarray}

\subsection{Matrix elements of the velocity operator}

With $\mathbf{v} 
= \partial H/\partial \mathbf{p}$ and the Hamiltonian (\ref{eq1}) the operators $v_x$ and $v_y$  read
\begin{equation}
v_x = \tau \upsilon_F\left(
\begin{array}
[c]{cc}%
\sigma_x  &  0\\
0  &  \sigma_x
\end{array}
\right)  , \label{eq22}%
\end{equation}
%
%
\begin{equation}
v_y = \upsilon_F\left(
\begin{array}
[c]{cc}%
\sigma_y  &  0\\
0  &  -\sigma_y
\end{array}
\right)  , \label{eq23}%
\end{equation}
with $\sigma_i$ $(i=x,y)$ 
the Pauli matrices for the two basis functions of the energy bands of a monolayer \cite{xiao12}. For the calculations we need the  matrix elements, 
$\langle \Psi_{\lambda \mu \mathbf{k}} \vert v_\nu \vert \Psi_{\lambda^{\prime} \mu^{\prime} \mathbf{k^\prime}} \rangle$, 
$\nu = x, y$ and $\Psi_{\lambda \mu \mathbf{k}}$  given in Eq.~(\ref{eq18}). All calculations are done for a specific valley and a specific spin state. We introduce the  notation 
\begin{equation}
\langle \Psi_{\alpha} \vert v_\nu \vert \Psi_{\alpha^{\prime}} \rangle = \langle \alpha \vert v_\nu \vert \alpha^{\prime} \rangle = v_{\nu, \alpha \alpha^{\prime}} , \label{eq25}%
\end{equation}
where $\alpha$, $\alpha^{\prime}$ denote collectively 
the quantum numbers $\lambda, \mu, \mathbf{k}$, i.e., $\vert \alpha \rangle = \vert \lambda \mu \mathbf{k} \rangle$ and $\vert \alpha^{\prime} \rangle = \vert \lambda^{\prime} \mu^{\prime} \mathbf{k^{\prime}} \rangle$. The matrix elements of $v_\nu$ are diagonal in the index $\mathbf{k}$, that is, 
\begin{equation}
v_{\nu, \alpha \alpha^{\prime}} = v_{\nu, \lambda \mu; \lambda^{\prime} \mu^{\prime}} ( \mathbf{k} ) \delta_{\mathbf{k k^{\prime}}} , \label{eq26}%
\end{equation}
with 
\begin{align}
\hspace{-0.7cm}
&v_{x, \lambda \mu; \lambda^{\prime} \mu^{\prime}} ( \mathbf{k} )   = M 
\Big[\Big( \frac{\epsilon_k \text{e}^{i \tau \varphi_{\mathbf{k}}}}{E_{\lambda \mu \mathbf{k}}-\delta_1} + \frac{\epsilon_k \text{e}^{- i \tau \varphi_{\mathbf{k}}}}{E_{\lambda^{\prime} \mu^{\prime} \mathbf{k}} -\delta_1} \Big)  \nonumber
\\*
&\hspace{-0.42cm}\times\eta_{\lambda \mu \mathbf{k}} \eta_{\lambda^{\prime} \mu^{\prime} \mathbf{k}} 
 + \frac{\epsilon_k \text{e}^{- i \tau \varphi_{\mathbf{k}}}}{E_{\lambda \mu \mathbf{k}}+\delta_2-\delta_\lambda} + \frac{\epsilon_k \text{e}^{ i \tau \varphi_{\mathbf{k}}}}{E_{\lambda^{\prime} \mu^{\prime} \mathbf{k}}+\delta_2-\delta_\lambda}
\Big] ,
\label{eq27}%
\end{align}
%
$M=N_{\lambda \mu \mathbf{k}} N_{\lambda^{\prime} \mu^{\prime} \mathbf{k}} v_F$, and 
%

\begin{align}
\hspace{-0.5cm}&v_{y, \lambda \mu; \lambda^{\prime} \mu^{\prime}} ( \mathbf{k} )   = M 
i \tau \Big[\Big( - \frac{\epsilon_k \text{e}^{i \tau \varphi_{\mathbf{k}}}}{E_{\lambda \mu \mathbf{k}}-\delta_1} + \frac{\epsilon_k \text{e}^{- i \tau \varphi_{\mathbf{k}}}}{E_{\lambda^{\prime} \mu^{\prime} \mathbf{k}} -\delta_1} \Big) \nonumber\\ 
&\hspace{-0.45cm}\times\eta_{\lambda \mu \mathbf{k}} \eta_{\lambda^{\prime} \mu^{\prime} \mathbf{k}}  
+ \frac{\epsilon_k \text{e}^{- i \tau \varphi_{\mathbf{k}}}}{E_{\lambda \mu \mathbf{k}}+\delta_2-\delta_\lambda} - \frac{\epsilon_k \text{e}^{ i \tau \varphi_{\mathbf{k}}}}{E_{\lambda^{\prime} \mu^{\prime} \mathbf{k}}+\delta_2-\delta_\lambda}
\Big] .
\label{eq29}%
\end{align}
Note also the equality
\begin{equation}
v_{\nu, \lambda^{\prime} \mu^{\prime}; \lambda \mu} ( \mathbf{k} ) =  v_{\nu, \lambda \mu; \lambda^{\prime} \mu^{\prime}}^\ast ( \mathbf{k} )   . \label{eq30}%
\end{equation}
due to 
the hermiticity of the velocity operators. The 
matrix elements (\ref{eq27}) and (\ref{eq29}) will be used in the evaluation of the Berry curvatures and magnetic moments.

\section{Berry curvature and orbital magnetic moment}

Below we calculate and discuss 
the Berry curvature and orbital magnetic moment of bilayer WSe$_2$ since 
they determine the behaviour of the OM
and that of the ANE.



\subsection{Berry curvature}


%
Recently there have been several works on the effects of symmetry breaking \cite{kim18} and tunability of the Berry curvature \cite{kormanyos18} with important consequences for the spin- and valley-Hall effects in monolayer and bilayer MoS$_2$. There are also important consequences for 
the OM, 
and  the ANE, 
 as we discuss below. The Berry curvature \cite{chang96} of a band labeled by $\lambda$ and $\mu$ is defined as
\begin{equation}
\mathbf{\Omega}_{\lambda \mu} ( \mathbf{k} ) = i \left\langle \nabla_{\mathbf{k}} u_{\lambda \mu} ( \mathbf{k} ) \vert \times \vert \nabla_{\mathbf{k}} u_{\lambda \mu} ( \mathbf{k} ) \right\rangle , \label{eq37}%
\end{equation}
where $\vert u_{\lambda \mu} ( \mathbf{k} ) \rangle = \sqrt{S} \text{e}^{-i \mathbf{k} \cdot \mathbf{r}} \vert \Psi_{\lambda \mu \mathbf{k}} \rangle$, and the valley and spin indices, $\tau$ and $s_z$, have been suppressed for clarity. It can be expressed in a gauge-invariant form that is more convenient for numerical computations. 
In 2D materials only the $k_z$-component survives and takes the form 
\begin{eqnarray}
\hspace*{-0.75cm}
\Omega_{\lambda \mu} ( \mathbf{k} ) = - 2 \hbar^2 \text{Im}\sum_{\substack{\lambda^{\prime} \mu^{\prime}\\
(\lambda^{\prime} \neq \lambda)}} \frac{\langle u_{\lambda \mu} \vert v_x  \vert u_{\lambda^{\prime} \mu^{\prime}} \rangle \langle u_{\lambda^{\prime} \mu^{\prime}} \vert v_y \vert u_{\lambda \mu} \rangle}{\left( E_{\lambda \mu \mathbf{k}} - E_{\lambda^{\prime} \mu^{\prime} \mathbf{k}} \right)^2} , 
\label{eq55}%
\end{eqnarray}
where $v_\nu = (1 / \hbar ) \nabla_{k_\nu} H$. Using 
the  matrix elements of $v_x$ and $v_y$,
 given by Eqs.~(\ref{eq27}) and (\ref{eq29}),   Eq.~(\ref{eq55})  becomes
\begin{eqnarray}
 \hspace*{-0.70cm}
\Omega_{\lambda \mu} ( \mathbf{k} ) = - 2 \hbar^2 \sum_{\substack{\lambda^{\prime} \mu^{\prime}\\
(\lambda^{\prime} \neq \lambda)}} \frac{\text{Im} \Big[ v_{x, \lambda \mu; \lambda^{\prime} \mu^{\prime}} ( \mathbf{k} ) v_{y, \lambda \mu; \lambda^{\prime} \mu^{\prime}}^{\ast} ( \mathbf{k} ) \Big]}{\left( E_{\lambda \mu \mathbf{k}} - E_{\lambda^{\prime} \mu^{\prime} \mathbf{k}} \right)^2} . 
\label{eq59}%
\end{eqnarray}
\begin{figure}[ptb]
\vspace{-0cm}
\begin{center}
\includegraphics[height=6.2cm, width=8.6cm ]{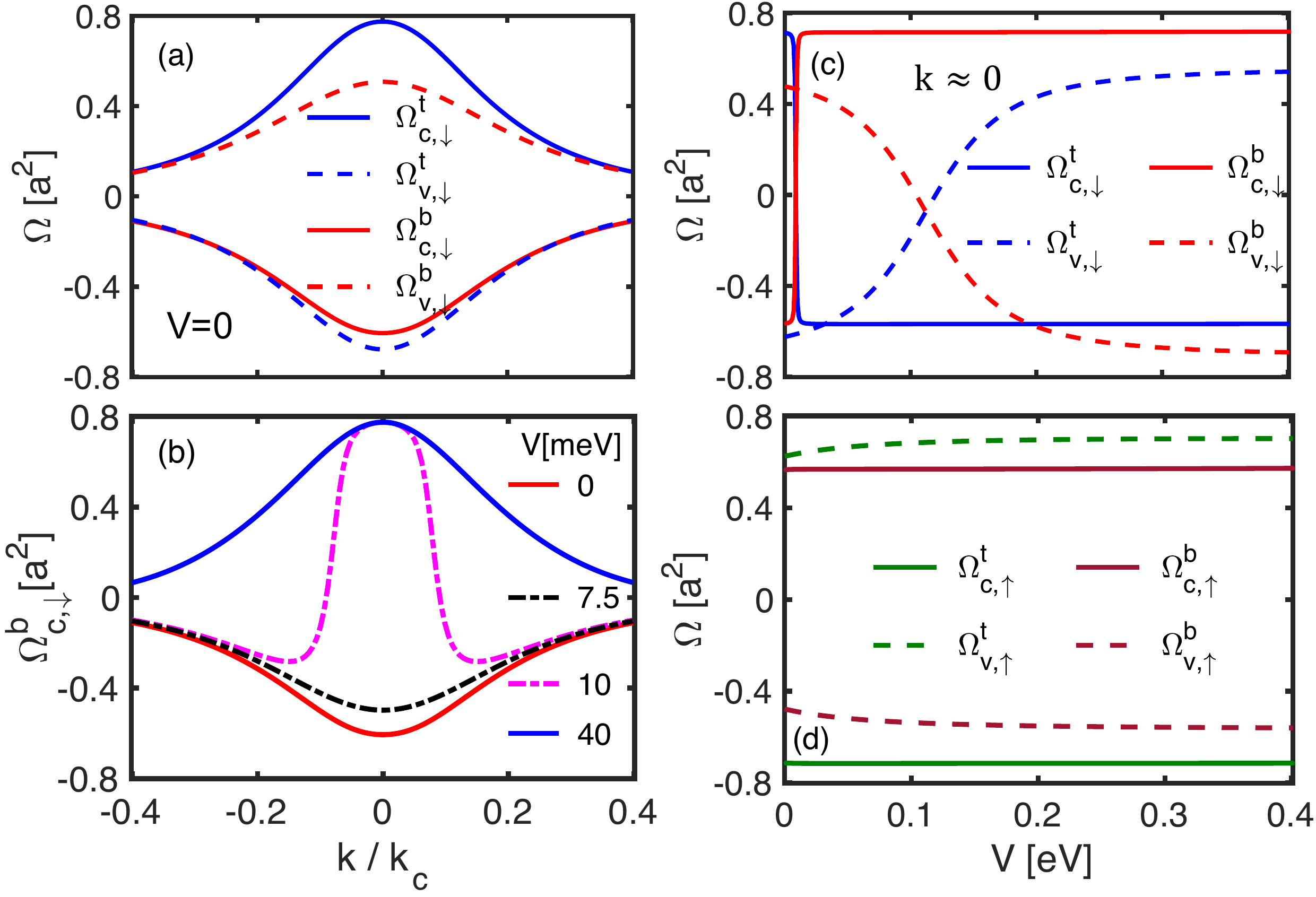}
\end{center}
\vspace*{-0.6cm} \caption{(Colour online) Berry curvature $\Omega ( \mathbf{k} )$ near the $K$ valley. Distributions of $\Omega ( \mathbf{k} )$ have opposite signs in the $K^{\prime} = - K$ valley. (a) Berry curvature of spin-down bands versus $k/k_c$ for zero electrostatic potential $(V = 0)$, and (b) $\Omega_{c,\downarrow}^b$ versus $k / k_c$ for increasing values of $V$. (c) Berry curvature of spin-down bands versus $V$ for $k/k_c \simeq 0$. (d) The same as in (c) but for spin-up bands. }
\label{fig:fig2}%
\end{figure}

The Berry curvature of a band arises due to the restriction to a single-band description, i.e., it can be regarded as the result of the ``residual" interaction of other adjacent bands. It becomes large when other bands are close, which is evident from Eq.~(\ref{eq55}), where a sum over all other bands is performed weighted by the inverse energy difference squared. In bilayer TMDCs the origin of the Berry curvature is the SOC and has an interesting effect; namely, it leads to finite Berry curvature even for $V=0$. This is because the SOC term, though it preserves $TR$ and $P$ symmetries, for a particular valley and spin state it appears in the diagonal matrix elements of the Hamiltonian Eq.~(\ref{eq1}) in the same way as the electric field does. This is shown in Fig.~2(a) where we plot the Berry curvature of spin-down bands for the $K$ valley and $V=0$ versus $k / k_c$, where $k_c = \pi / a$ and $a$ is the lattice constant. The values of the parameters are the same as those in Fig.~1 and we use $c (v)$ for the conduction (valence) band and $t (b)$ for the top (bottom) layer. It can be seen that the Berry curvature of a conduction/valence band is strongly enhanced and peaked at the valley extrema. The asymmetry of the Berry curvatures of the bands with respect to $k$-axis is a consequence of the broken particle-hole symmetry which originates from the unequal values of the SOC for electrons and holes, and also because the interlayer coupling for electrons vanishes. Due to the preservation of both $TR$ and $P$ symmetries, the 
Berry curvatures for  spin-up bands are just the reverse of those for spin-down bands, i.e., $\Omega_{n, \uparrow}^{\ell} ( \mathbf{k} ) = - \Omega_{n, \downarrow}^{\ell} ( \mathbf{k} )$, where $n = c ( v )$ and $\ell = t ( b )$. Moreover, at the $K^{\prime} = - K$ valley (not shown) the Berry curvature has the same magnitude but opposite sign, as required by $TR$ symmetry. 

In Fig.~2(b) we illustrate the effect of the electric field, which can tune the Berry curvature. We only show the conduction-band Berry curvature of the bottom layer $\Omega_{c, \downarrow}^{b}$. As $V$ increases from zero, $\Omega_{c, \downarrow}^{b}$ becomes more concentrated around the $K$ valley and when $V \simeq 7.5$meV it becomes small and reverses polarity thereafter (purple and blue lines for $10$ and $40$ meV). For higher $V$, it becomes positive and considerably larger compared to the $V = 0$ case. At the same time, the conduction-band Berry curvature of the top layer, $\Omega_{c, \downarrow}^{t}$, also decreases in magnitude and reverses polarity at the same value of $V$. This is shown in Fig.~2(c) where we plot the Berry curvature versus the 
potential $V$ for $k / k_c \simeq 0$. For $V \simeq 7.5$meV, which is the value of SOC for electrons, $\Omega_{c, \downarrow}^{b} = \Omega_{c, \downarrow}^{t}$. This is also the case for $\Omega_{v, \downarrow}^{b}$ and $\Omega_{v, \downarrow}^{t}$. However, this time the polarity inversion occurs at $V \simeq 112.5$meV, which is the value of SOC for holes. This is because 
the 
electric field turns the SOC effectively off for the spin-down electrons/holes as $V \rightarrow \lambda_c  / \lambda_v$. This is clearly seen in the diagonal elements of the Hamiltonian (\ref{eq1}) where, for the $K$ valley and spin-down electrons, the expressions $- \lambda_{c(v)} +V$ for the top layer and $\lambda_{c(v)} - V$ for the bottom layer become zero once $V$ reaches $\lambda_c$ or $\lambda_v$. The simultaneous polarity inversion of the Berry curvatures that occurs at $\lambda_c$ and $\lambda_v$ should rather be expected because the Chern number is a topological invariant. Therefore, a peak inversion of the Berry curvature of a band must be accompanied by an opposite peak inversion of the Berry curvature of another band such that the Chern number (summed over valleys and spins) retains its initial value.

On the other hand, the effect of the electric field on the Berry curvature for spin-up bands is much weaker, as shown in Fig.~2(d) where we plot the Berry curvatures versus $V$. At the $K^{\prime}$ valley, the polarity inversion occurs for  spin-up electrons but not for the spin-down ones. 

\subsubsection{Bilayer vs monolayers.}

A better understanding of the Berry curvature properties is reached by contrasting a bilayer WSe$_2$ 
with two decoupled monolayers. In the latter case, 
the Hamiltonian (\ref{eq1}) is block diagonal ($\gamma=0$) and we can find the eigenfunctions and eigenvalues for each layer separately. The eigenfunctions of the top layer are given by
\begin{equation}
\Psi_{\lambda \mathbf{k}} ( \mathbf{r} ) = \frac{1}{
[\left( E_{\lambda \mathbf{k}} - \delta_1 \right)^2 +
\epsilon_k^2]^{1/2}
} \left(
\begin{array}
[c]{c}%
\vspace{0.07in} \epsilon_k \tau \text{e}^{-i \tau \varphi_{\mathbf{k}}}\\
E_{\lambda \mathbf{k}} - \delta_1%
\end{array}
\right)\label{eq1000}%
\end{equation}
and the eigenvalues by
\begin{equation}
E_{\lambda \mathbf{k}} = 
(\delta_1 + \delta_2)/2 + (\lambda/2) 
[(\delta_1 - \delta_2)^2 +4\epsilon_k^2]^{1/2}, 
\label{eq1001}%
\end{equation}
where 
the valley and spin indices $\tau$ and $s_z$ have been suppressed. The corresponding eigenfunctions and eigenvalues  for the bottom layer are obtained from those of the top layer by taking the complex conjugate and making the replacement $\delta_1 \rightarrow -\delta_2+\delta_\lambda$ and $\delta_2 \rightarrow -\delta_1-\delta_\lambda$. The Berry curvature for the top layer is obtained from Eq.~(\ref{eq55}) with $\mu=\mu'=\pm1$.
The evaluation proceeds as in 
the bilayer case and gives the analytic result
\begin{equation}
\Omega_{c}^{t} ( \mathbf{k} ) = - \tau  \frac{2 \hbar^2 v_F^2 ( \delta_1 - \delta_2
 )}{ \left \{ (\delta_1 -\delta_2 
 )^2 + 4 \epsilon_k^2\right \}^{3/2}}
\label{eq10031}%
\end{equation}
for the spin-dependent Berry curvature of the conduction band, and $\Omega_{v}^{t} ( \mathbf{k} ) = - \Omega_{c}^{t} ( \mathbf{k} )$ for the valence band. The Berry curvature for the bottom layer is obtained from Eq.~(\ref{eq10031}) with the replacements $\delta_1 \rightarrow -\delta_2+\delta_\lambda$ and $\delta_2 \rightarrow -\delta_1-\delta_\lambda$. Importantly, the electric field has dropped out and thus it has no effect.
\begin{figure}[ptb]
\begin{center}
\hspace*{-0.5cm}
\includegraphics[
height=11cm, width=8cm ]{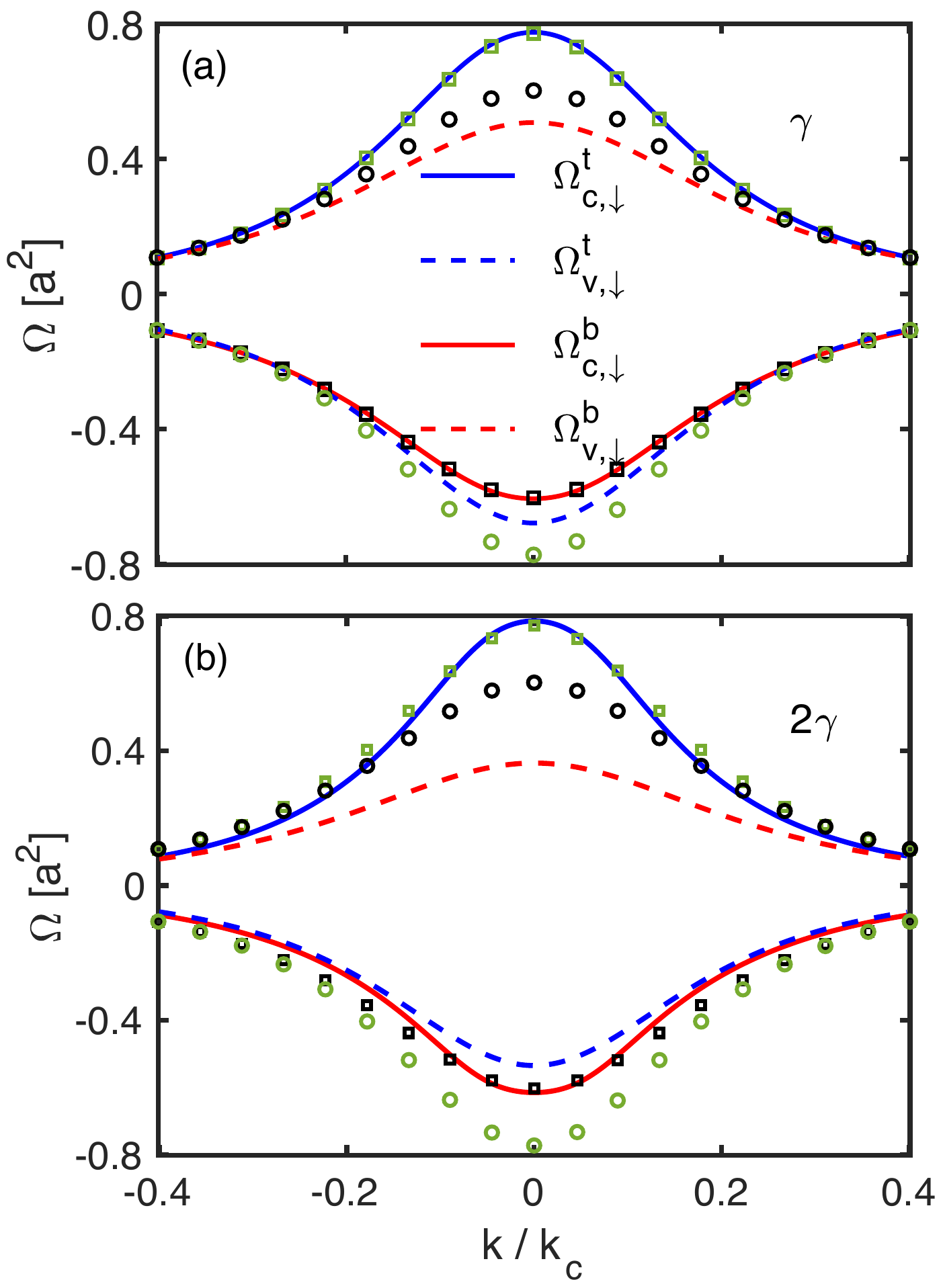}
\end{center}
\vspace*{-0.6cm} \caption{(Colour online) (a) Comparison of $\Omega ( \mathbf{k} )$ near the $K$ valley for the bilayer (solid and dashed curves, see Fig.~2(a)) with that for two monolayers. In the latter case the $\square$ symbols show the results for conduction bands, and the $\bigcirc$ symbols the results for valence bands. Green color corresponds to bands in the top layer, black color to bands in the bottom layer. (b) The same as in (a) but for interlayer coupling twice as large, $2\gamma = 0.134$ meV.}
\label{fig:fig2}%
\end{figure}

In Fig.~3(a) we compare the Berry curvature of the spin-down bands for the bilayer (solid and dashed lines) with that for the monolayers. We used the same parameter values as in Fig.~2(a). The Berry curvature $\Omega_{c (v), \downarrow} ^{t(b)}( \mathbf{k} )$ for the decoupled layers is shown by using $\square$ for conduction bands, and $\bigcirc$ for valence bands. Green color corresponds to the top layer and black color to the bottom layer. It is seen that the Berry curvatures of the conduction bands in either the  top or bottom layer of the bilayer WSe$_2$ are almost identical to those of the decoupled monolayers. This is due to the weak SOC for electrons. However, the strong SOC for holes combined with the interlayer coupling $\gamma$ causes the Berry curvature of the valence bands of both layers to decrease.  

It is instructive to make the same comparison when the interlayer coupling becomes twice as large, i.e., $\gamma \rightarrow 2 \gamma$. This is shown in Fig.~3(b). In this case the Berry curvatures of the valence bands of the bilayer are substantially different from those of the decoupled monolayers. The gradually smaller magnitude of the Berry curvature of a band is due to the gradually smaller influence of other adjacent bands; their separation increases as $\gamma$ increases (see discussion in Sec.~IIIA). We remark that if SOC is neglected, the Berry curvatures of both valence and conduction bands of the bilayer vanish. However, in the decoupled monolayers the Berry curvature is nonzero, as is evident from Eq.~(\ref{eq10031}). We also notice that the Berry curvatures of the decoupled layers are different from each other. The physical origin of this lies in the different signs \cite{gong13} of the SOC in the two layers within a given valley. This is a consequence of the fact that the two layers are rotated by 180$^{\circ}$ with respect to each other.
\begin{figure}[ptb]
\hspace*{-0.2cm}
\begin{center}
\includegraphics[
height=6.2cm, width=8.6cm ]{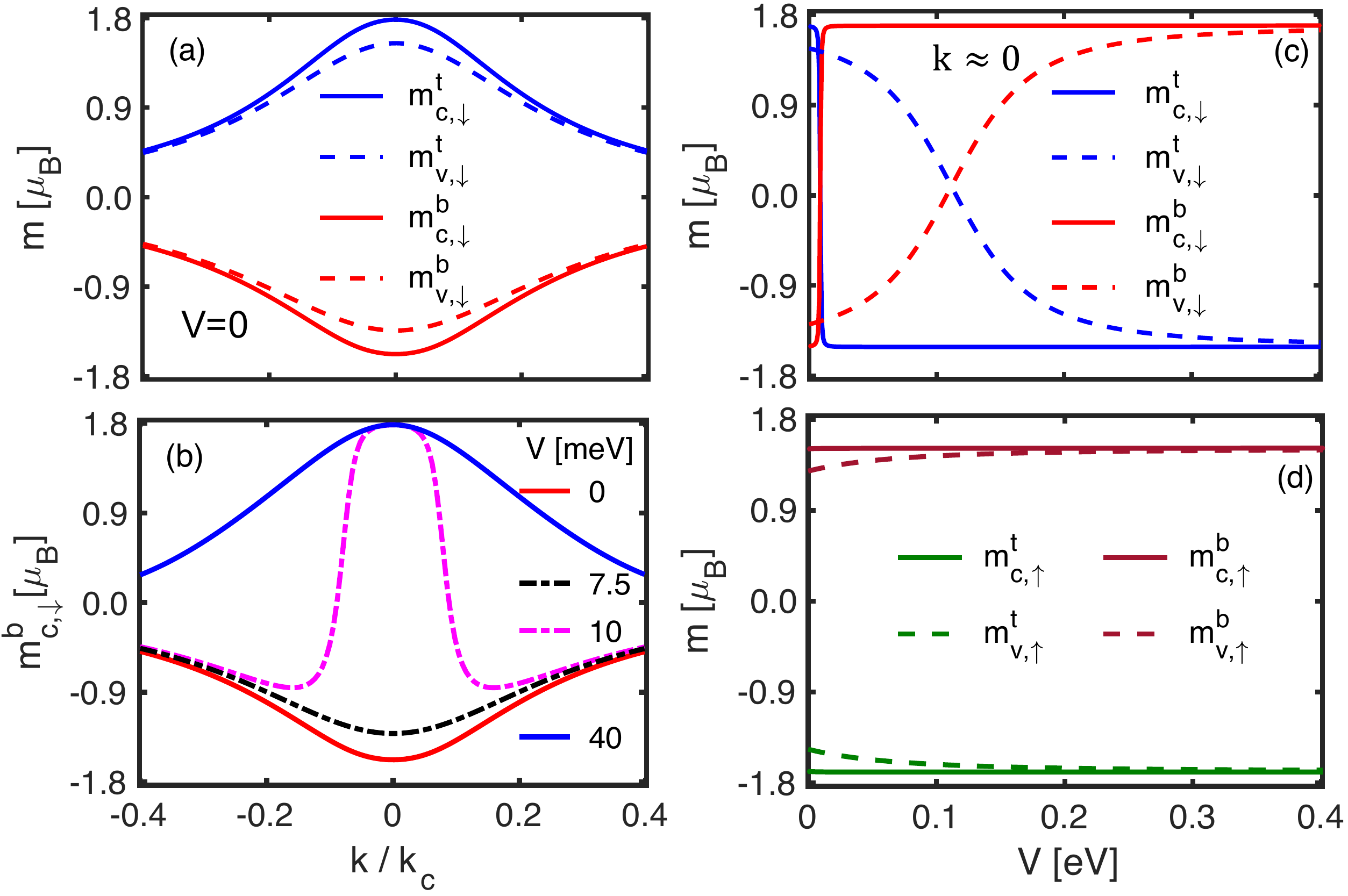}
\end{center}
\vspace*{-0.55cm} \caption{(Colour online) Orbital magnetic moment $m ( \mathbf{k} )$ near the $K$ valley in units of Bohr magnetons $\mu_B = e \hbar / 2 m$. $m ( \mathbf{k} )$ has distribution similar to that of $\Omega ( \mathbf{k} )$ (see Fig.~2). (a) $m ( \mathbf{k} )$ of spin-down bands versus $k / k_c$ and $V = 0$. (b) $m_{c, \downarrow}^{b}$ versus $k / k_c$ for increasing values of $V$. (c) $m ( \mathbf{k} )$ of spin-down bands versus $V$ for $k / k_c \simeq 0$. (d) The same as in (c) but for spin-up bands. }%
\label{fig:fig3}%
\end{figure}

\subsection{Orbital magnetic moment}

The orbital magnetic moment of Bloch electrons is given by \cite{chang96}
\begin{eqnarray}
\nonumber \hspace*{-0.0cm}
\mathbf{m}_{\lambda \mu} ( \mathbf{k} ) = \frac{e}{2 i \hbar} \langle \nabla_{\mathbf{k}} u_{\lambda \mu} ( \mathbf{k} ) \vert \times \left[ H ( \mathbf{k} ) - E_{\lambda \mu \mathbf{k}} \right] \vert \nabla_{\mathbf{k}} u_{\lambda \mu} ( \mathbf{k} ) \rangle .\\*
\label{eq64}%
\end{eqnarray}
It originates from the self-rotation of the electron wave packet around its center of mass \cite{xiao06} and therefore carries orbital angular momentum in addition to its spin angular momentum. 
Using the identity 
\begin{equation}
\langle \nabla_{\mathbf{k}} u_{\lambda \mu} \vert u_{\lambda^{\prime} \mu^{\prime}} \rangle = \frac{\langle u_{\lambda \mu} \vert \nabla_{\mathbf{k}} H \vert u_{\lambda^{\prime} \mu^{\prime}} \rangle}{E_{\lambda \mu \mathbf{k}} - E_{\lambda^{\prime} \mu^{\prime} \mathbf{k}}} 
\label{eq65}%
\end{equation}
and inserting the unity operator $1 = \sum_{\lambda^{\prime} \mu^{\prime}} \vert u_{\lambda^{\prime} \mu^{\prime}} \rangle \langle u_{\lambda^{\prime} \mu^{\prime}} \vert$ we can rewrite the $k_z$ component of Eq.~(\ref{eq64}) as
%
\begin{eqnarray}
 \hspace*{-0.6cm}
m_{\lambda \mu} ( \mathbf{k} ) = - \hbar e \sum_{\substack{\lambda^{\prime} \mu^{\prime}\\
(\lambda^{\prime} \neq \lambda)}} \frac{\text{Im} \Big[ v_{x, \lambda \mu; \lambda^{\prime} \mu^{\prime}} ( \mathbf{k} ) v_{y, \lambda \mu; \lambda^{\prime} \mu^{\prime}}^{\ast} ( \mathbf{k} ) \Big]}{ E_{\lambda \mu \mathbf{k}} - E_{\lambda^{\prime} \mu^{\prime} \mathbf{k}} } . 
\label{eq68}%
\end{eqnarray}
The orbital magnetic moment of spin-down bands near the $K$ valley is plotted in Fig.~4(a) versus $k / k_c$ for $V = 0$. Its distribution is similar to that of $\Omega ( \mathbf{k} )$. However, in contrast to $\Omega ( \mathbf{k} )$, we notice that the magnetic moments of both conduction and valence bands in the top layer are positive while in the bottom layer they are negative. The effect of the electric field on the magnetic moments is shown in Fig.~4(b). We only show the conduction-band magnetic moment, $m_{c, \downarrow}^{b}$, of the bottom layer. We notice that, as $V$ increases from zero, $m_{c, \downarrow}^{b}$ changes sign at $V \simeq \lambda_c$. This behavior is the same as that for the Berry curvature, see Fig.~2(b). In Fig.~4(c) we plot the magnetic moment versus the potential $V$ for $k / k_c \simeq 0$. We notice that increasing the electric field from zero causes inversion of the magnetic moments of the conduction/valence bands in the top and bottom layers as $V \rightarrow \lambda_c / \lambda_v$. The reason is the same as that for $\Omega ( \mathbf{k} )$. 

The orbital magnetic moment of spin-up bands is barely affected by the electric field, see Fig.~4(d).

\begin{figure}[ptb]
\vspace*{0.25cm}
\begin{center}
\includegraphics[height=3.4cm, width=8.6cm ]{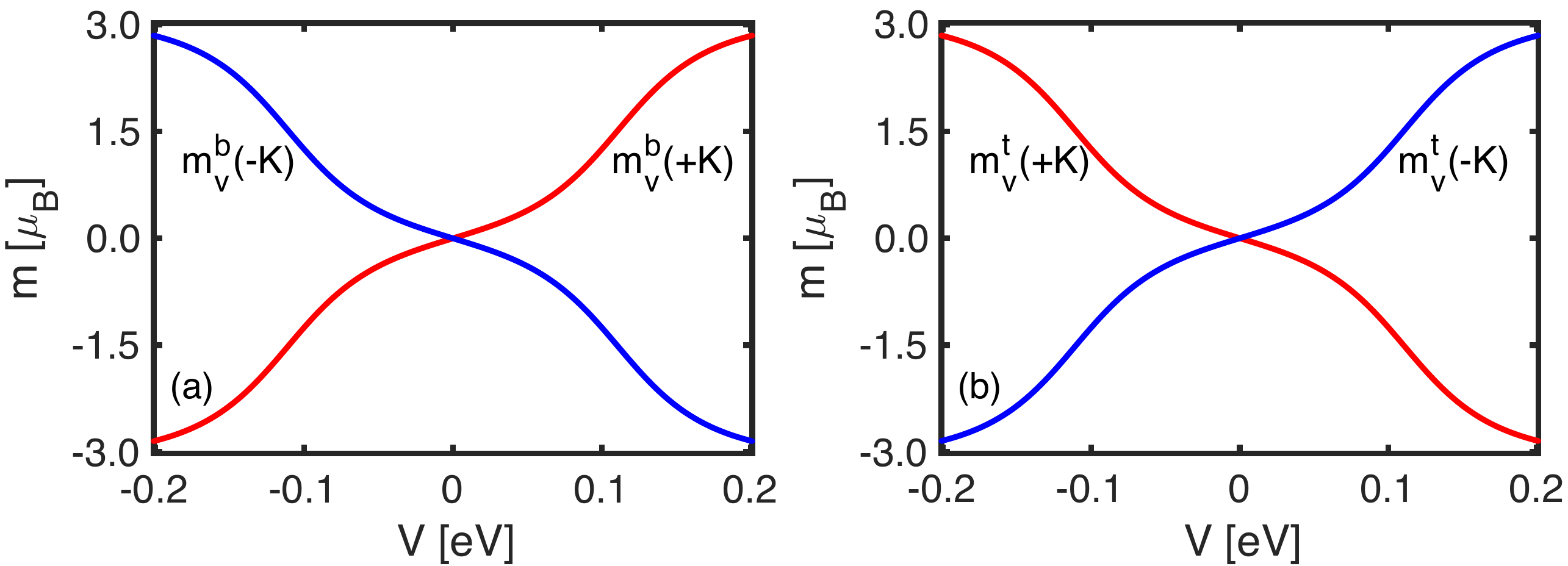}
\end{center}
\vspace*{-0.55cm} \caption{(Colour online) (a) Orbital magnetic moment at $\pm K$, $m_v^b ( \pm K ) = m_{v, \downarrow}^b + m_{v, \uparrow}^b$, of the valence band in the bottom layer versus electrostatic potential $V$. (b) The same as in (a) but for the top layer.} 
\label{fig:fig1}%
\end{figure}
A perpendicular electric field offers the possibility of switching on/off and tuning the Berry curvature and orbital magnetic moment near the Dirac valleys. This is illustrated in Fig.~5(a) where we plot the orbital magnetic moment at $\pm K$ valleys of the valence band in the bottom layer, $m_v^b ( \pm K ) = m_{v, \downarrow}^b + m_{v, \uparrow}^b$. Figure 5(b) is for the top layer, $m_v^t ( \pm K ) = m_{v, \downarrow}^t + m_{v, \uparrow}^t$. {At $V = 0$, the combined effect of the $TR$ and $P$ symmetries causes vanishing of the magnetic moment. For nonzero values of $V$, the $P$ symmetry is broken and consequently nonzero $m_v^{b ( t )} ( \pm K )$ appears near the band edge. In the top layer the magnetic moments have opposite signs from those in the bottom layer, which has important consequences for the OM, as discussed in Sec.~IV. We also notice that $\mathbf{ m }$ is an odd function of the electrostatic potential. Hence, electrical control of valley-contrasting magnetic moment and Berry curvature, suggests the possibility of manipulating topological quantum phenomena in bilayer WSe$_2$ and other similar TMDCs. The results shown in Fig.~5 are very similar to those obtained with density functional theory (DFT) calculations for a bilayer MoS$_2$ \cite{wu13}. For electric field $E = 10$ mV$/\text{\AA}$, corresponding to $V \simeq 70$ meV, we estimate $m = 0.25$ $\mu_B$ for the bilayer WSe$_2$, which is slightly larger than the DFT result of $m \simeq 0.2$ $\mu_B$ for the bilayer MoS$_2$.


%

\section{Orbital Magnetization}

The modern theory of OM 
\cite{xiao05,thonhauser05} focuses on a crystalline system of independent Bloch electrons in the presence of $TR$ symmetry breaking. 
In this theory, 
the OM 
originates from the orbital magnetic moment of carriers and from a Berry curvature correction \cite{xiao07,xu14}. It has been studied in various systems including TMDCs \cite{tahir14}, topological insulators \cite{tahir15}, systems with arbitrary band topology \cite{nouraf14}, and more recently in Weyl semimetals \cite{nouraf19}.

On the other hand, in systems with $TR$ symmetry but with broken $P$ symmetry, the intrinsic magnetic moment associated with the valley pseudospin is analogous to the Bohr magneton which is associated with the electron spin. In this context, the OM 
is more appropriately called valley magnetization, and can be used in practical applications; for example, in 2D materials with broken $P$ symmetry, a population difference in the two valleys may be detected as a signal of OM 
\cite{xiao07} whose sign is different for $K$ and $K^{\prime}$ electrons.

The free energy 
$F$ in 
a weak magnetic field $\mathbf{B}$ is
\begin{equation}
F = - \frac{1}{\beta} \sum_{\lambda \mu \mathbf{k}} \ln \left [ 1 + \text{e}^{- \beta ( E_M - E_F )} \right ] , 
\label{eq291}%
\end{equation}
where the electron energy $E_M = E_{\lambda \mu \mathbf{k}} - \mathbf{m} (\mathbf{k}) \cdot \mathbf{B}$ includes a correction due to the orbital magnetic moment $\mathbf{m} ( \mathbf{k} )$. Further, $\beta = 1 / k_B T$ with $k_B$ the Boltzmann constant, $T$ is the temperature, and $E_F$ is the chemical potential. 
The OM 
of a band is given by $\mathbf{M}_{\lambda \mu} = - \left( \partial F / \partial \mathbf{B} \right)_{E_F, T} = 
\mathbf{M}_{\lambda \mu}^{(o)} + \mathbf{M}_{\lambda \mu}^{(b)}$ where
\begin{eqnarray}
\nonumber
\mathbf{M}_{\lambda \mu}^{(o)} =  \frac{a^2}{4 \pi^2} \int d^2 k \left[ f_{\lambda \mu}^{\uparrow} ( \mathbf{k} ) \mathbf{m}_{\lambda \mu}^{\uparrow} ( \mathbf{k} ) + f_{\lambda \mu}^{\downarrow} ( \mathbf{k} ) \mathbf{m}_{\lambda \mu}^{\downarrow} ( \mathbf{k} ) \right] , \\*
\label{eq292}%
\end{eqnarray}
\begin{eqnarray}
\nonumber 
\hspace*{-0.8cm}
\mathbf{M}_{\lambda \mu}^{(b)} = \frac{e a^2}{2 \pi  \beta h} \int d^2 k \Big\{ \mathbf{\Omega}_{\lambda \mu}^{\uparrow} ( \mathbf{k} ) \ln \Big[ 1 + \text{e}^{- \beta (E_{\lambda \mu \mathbf{k}}^{\uparrow} - E_F )} \Big] \\
+ \mathbf{\Omega}_{\lambda \mu}^{\downarrow} ( \mathbf{k} ) \ln \Big[ 1 + \text{e}^{- \beta (E_{\lambda \mu \mathbf{k}}^{\downarrow} - E_F )} \Big] \Big\}, 
\label{eq293}%
\end{eqnarray}
Equation (\ref{eq292}) is due to the thermodynamic average of the orbital magnetic moment,  $f_{\lambda \mu}^{\uparrow ( \downarrow )} ( \mathbf{k} )$ is the Fermi function, and $a$ the lattice constant. Equation (\ref{eq293}) %
is due to the center-of-mass motion of the wave packet and results from the Berry phase correction to the electron density of states \cite{xiao05}.
\begin{figure}[ptb]
\hspace*{-0.1cm}
\begin{center}
\includegraphics[
height=6.7cm, width=8.5cm ]{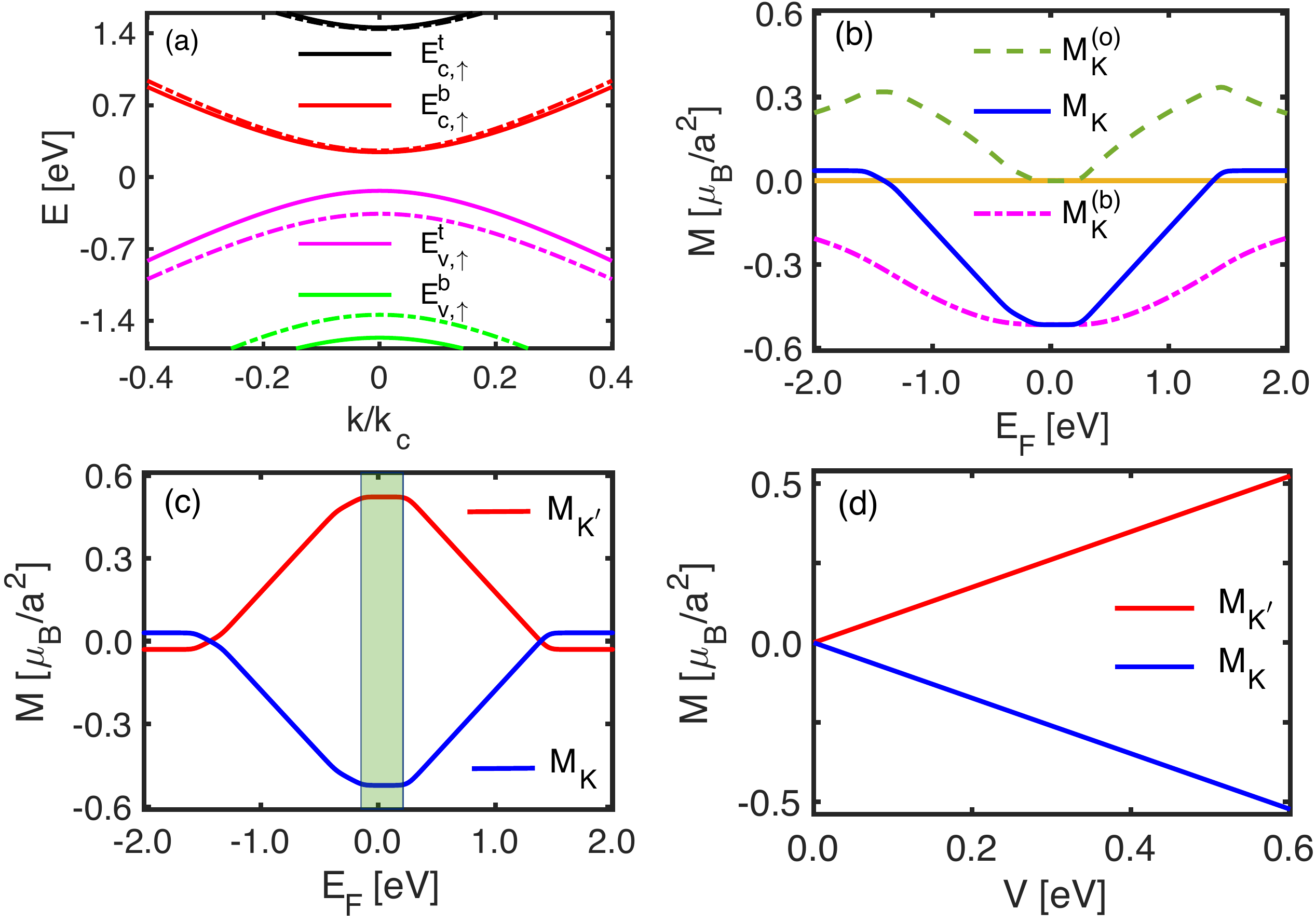}
\end{center}
\vspace*{-0.6cm} \caption{(Colour online) (a) Energy bands for $V=0.6$ eV, and  (b) Valley magnetization $M_K$ vs chemical potential $E_F$, with its two contributions $M_K^{(o)}$ and $M_K^{(b)}$. (c) Valley magnetization for $K$ and $K^{\prime}$ valleys. The shaded area is the gap. (d) Valley magnetizations $M_K$ and $M_{K^{\prime}}$ versus electrostatic potential $V$ for $E_F = 0$.   }
\label{fig:fig2}%
\end{figure}

The OM 
for the $K$ valley, $\mathbf{M}_{K} = \sum_{\lambda,\mu} \mathbf{M}_{\lambda \mu}$, is plotted in Fig.~6(b) versus the chemical potential $E_F$ for $V = 0$ (brown solid line) and for $V = 0.6$ eV (blue solid line). The temperature is $T = 300$ K. The energy bands for $V = 0.6$ eV are shown in Fig.~6(a) where solid and dashed lines represent spin-up and spin-down bands. The individual contributions $\mathbf{M}_{K}^{(o)} = \sum_{\lambda,\mu} \mathbf{M}_{\lambda \mu}^{(o)}$ and $\mathbf{M}_{K}^{(b)} = \sum_{\lambda,\mu} \mathbf{M}_{\lambda \mu}^{(b)}$ are also shown by the green dashed and magenta dashed-dotted lines, respectively. For $V = 0$, the $TR$ and $P$ symmetries enforce spin degeneracy of each band, so $f_{\lambda \mu}^{\uparrow} ( \mathbf{k} ) = f_{\lambda \mu}^{\downarrow} ( \mathbf{k} )$. On the other hand, these symmetries require $\mathbf{m}_{\lambda \mu}^{\uparrow} ( \mathbf{k} ) = - \mathbf{m}_{\lambda \mu}^{\downarrow} ( \mathbf{k} )$ and $\mathbf{\Omega}_{\lambda \mu}^{\uparrow} ( \mathbf{k} ) = - \mathbf{\Omega}_{\lambda \mu}^{\downarrow} ( \mathbf{k} )$ and therefore the OM vanishes. For $V = 0.6$ eV, we notice that for $E_F$ below the edge of the lowest valence band (green line in Fig.~6(a)) the valley magnetization $\mathbf{M}_{K}$ is seen to be constant. This is due to the fact that $\Omega_{v, \uparrow ( \downarrow )}^{t}$ have opposite signs from $\Omega_{v, \uparrow ( \downarrow )}^{b}$ as shown in Figs.~2(c) and 2(d). This is also the case for the magnetic moments as shown in Figs.~3(c) and 3(d). This means that the valence bands $E_{v, \uparrow ( \downarrow )}^{t}$ and $E_{v, \uparrow ( \downarrow )}^{b}$ carry opposite-circulating currents that compete with each other giving rise to opposite contributions to the valley magnetization. The small excess amount of positive circulating current over that with negative circulation leads to the small positive value of $\mathbf{M}_{K}$. After $E_F$ 
crosses the edge of the lower valence band the magnetization increases in absolute value because the circulating currents of the higher valence band are left unbalanced. The valley magnetization remains constant as $E_F$ scans the insulating gap. This is consistent with the fact that $\mathbf{M}$ changes linearly, when $E_F$ is varied in the gap, only if the Chern number is nonzero, and remains constant otherwise. This is embodied in the relation $d \mathbf{M} / d E_F = \left( e / h \right) \mathcal{C}$ \cite{ceresoli06,tabert15}, where $\mathcal{C}$ is the Chern number. We have numerically verified  that the Chern number for each valley (summed over spin) vanishes, as expected for a band insulator. Upon further increase of $E_F$ 
the magnetization exhibits a symmetrical behaviour as a function of it, as the circulating currents of $E_{c, \uparrow ( \downarrow )}^{b}$ are unbalanced until $E_F$ crosses the bottom of the higher conduction band $E_{c, \uparrow ( \downarrow )}^{t}$. 

The magnetization $\mathbf{M}_{K^{\prime}}$ for the $K^{\prime}$ valley is equal and opposite to $\mathbf{M}_{K}$, as required by $TR$ symmetry [see Fig.~6(c)] so that the total magnetization vanishes. The shaded area shows the band gap. In Fig.~6(d) we show that the valley magnetizations $\mathbf{M}_{K}$ and $\mathbf{M}_{K^{\prime}}$ change linearly with increasing 
potential $V$ for $E_F$ in the gap. Note that in the presence of $P$ symmetry at $V = 0$, $\mathbf{M}_{K}$ and $\mathbf{M}_{K^{\prime}}$ vanish.  

\section{Anomalous Nernst effect}

Conventionally, the Nernst effect \cite{ziman01} occurs in the presence of a longitudinal temperature gradient and an external magnetic field, which provides a transverse velocity to the electrons by the Lorentz force. This leads to the generation of a transverse electric field. However, a nontrivial Berry curvature $\mathbf{\Omega} ( \mathbf{k} )$ of the bands can also give rise to Hall and Nernst responses in each valley as a consequence of an anomalous velocity term generated by $\mathbf{\Omega} ( \mathbf{k} )$, leading to valley ANE and AHE. In this respect, the ANE is the thermoelectric counterpart of the AHE. 

The anomalous Nernst response can be obtained using the semiclassical wave packet methods taking into account the OM 
of the carriers arising from the finite spread of the wave function \cite{xiao06}. In this approach, an intrinsic Hall current results when a temperature gradient is present, $j_x = \alpha_{xy} ( - \nabla_y T )$, from which a spin- and valley-dependent anomalous Nernst coefficient (ANC) can be extracted as
\begin{equation}
\alpha_{xy}^{\tau s_z} = \frac{e k_B}{\hbar} \sum_{\lambda \mu} \int \frac{d^2 k}{( 2 \pi )^2} \Omega_{\lambda \mu}^{\tau s_z} ( \mathbf{k} )  S_{\lambda \mu}^{\tau s_z} ( \mathbf{k} )  ,
\label{eq294}%
\end{equation}
where $S_{\lambda \mu}^{\tau s_z} ( \mathbf{k} ) = - f ( \mathbf{k} ) \ln f ( \mathbf{k} ) - ( 1 - f ( \mathbf{k} ) ) \ln ( 1 - f ( \mathbf{k} ) )$ is the entropy density and $f ( \mathbf{k} )$  the Fermi distribution function. In contrast to the anomalous Hall conductivity $\sigma_{xy}$, which depends only on the Berry curvature of the filled bands, the anomalous Nernst coefficient $\alpha_{xy}$ is a Fermi surface quantity, because $S_{\lambda \mu}^{\tau s_z} ( \mathbf{k} )$ vanishes for completely filled and completely empty bands. The ANC for a specific valley reads
\begin{eqnarray}
\hspace*{-0.7cm}
\alpha_{xy}^{\text{v}} = \alpha_0 \sum_{\lambda \mu} \int d^2 k \Big[ \Omega_{\lambda \mu}^{\uparrow} ( \mathbf{k} )  S_{\lambda \mu}^{\uparrow} ( \mathbf{k} ) + \Omega_{\lambda \mu}^{\downarrow} ( \mathbf{k} )  S_{\lambda \mu}^{\downarrow} ( \mathbf{k} ) \Big], 
\label{eq295}%
\end{eqnarray}
with $\alpha_0 = e k_B/4 \pi^2 \hbar$ and the integrand  evaluated at $\tau = \pm 1$. The anomalous spin Nernst coefficient 
(SNC) is 
\begin{eqnarray}
\hspace*{-0.7cm}
\alpha_{xy}^{\text{s}} = \alpha_0^{s} \sum_{\lambda \mu} \int d^2 k \Big[ \Omega_{\lambda \mu}^{\uparrow} ( \mathbf{k} )  S_{\lambda \mu}^{\uparrow} ( \mathbf{k} ) - \Omega_{\lambda \mu}^{\downarrow} ( \mathbf{k} )  S_{\lambda \mu}^{\downarrow} ( \mathbf{k} ) \Big], 
\label{eq296}%
\end{eqnarray}
where $\alpha_0^{s} = \alpha_0 \hbar/2e = k_B/8 \pi^2$. We have multiplied by $\hbar/2e$ to conform with the definition of a spin current \cite{kane05}.
\begin{figure}[t]
\vspace*{0.2cm}
\begin{center}
\includegraphics[height=3.5cm, width=8.6cm ]{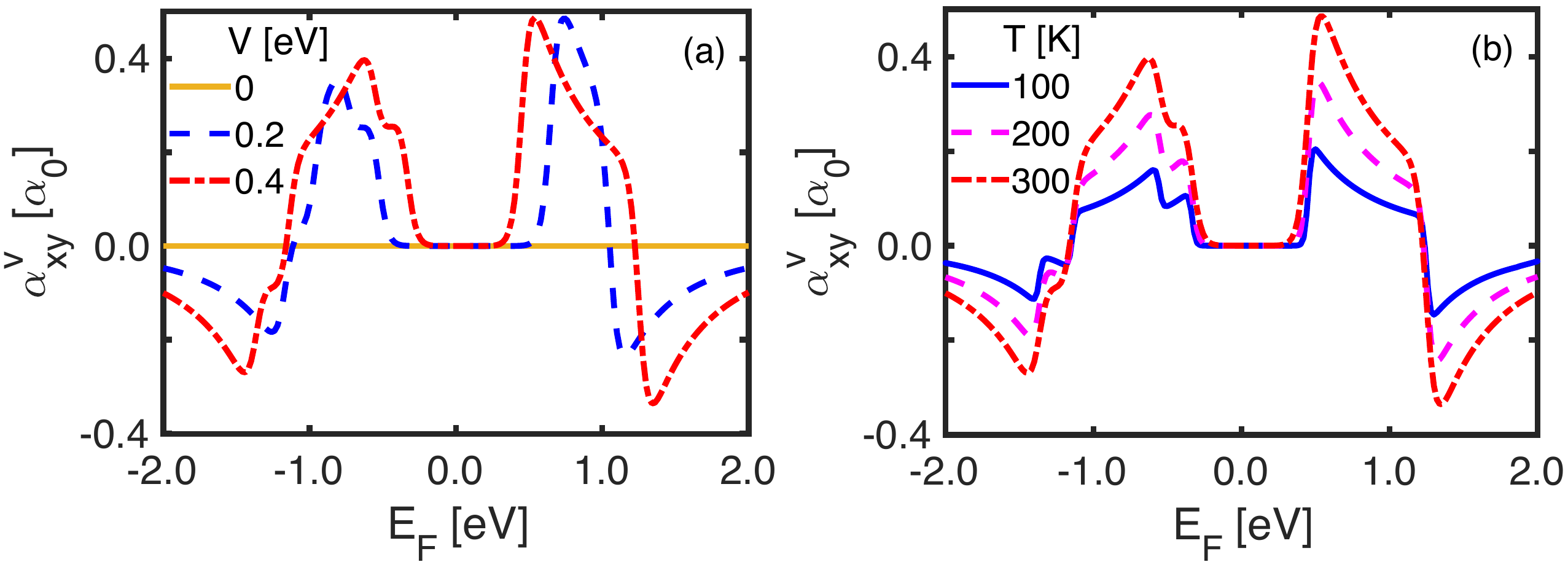}
\end{center}
\vspace*{-0.65cm} \caption{(Colour online) (a) ANC $\alpha_{xy}^{\text{v}}$ for the $K$ valley versus 
$E_F$ for increasing values of electrostatic potential $V$, in units $\alpha_0 = e k_B/4 \pi^2 \hbar$. The temperature is $T = 300$ K. (b) The same as in (a) for various temperatures and fixed $V = 0.4$ eV. }
\label{fig:fig1}%
\end{figure} 

The valley ANC is calculated from Eq.~(\ref{eq295}) by integrating the Berry curvature up to $k_F$. 
In Fig.~7(a) we show the ANC for the $K$ valley versus chemical potential for increasing values of $V$ at $T = 300$ K. For $V = 0$, inversion symmetry is preserved and therefore each band is spin-degenerate. This implies $S_{\lambda \mu}^{\uparrow} ( \mathbf{k} ) = S_{\lambda \mu}^{\downarrow} ( \mathbf{k} )$. On the other hand, as noted earlier, $\Omega_{\lambda \mu}^{\uparrow} ( \mathbf{k} ) = - \Omega_{\lambda \mu}^{\downarrow} ( \mathbf{k} )$ and therefore $\alpha_{xy}^{\text{v}}$ vanishes in this limit. For $V > 0$, the inversion symmetry is broken and the spin-degeneracy is lifted, leading to finite $\alpha_{xy}^{\text{v}}$. The ANC exhibits dip (peak) features as $E_F$ 
crosses the top of the lowest (highest) valence band. Their magnitudes and signs are proportional to those of the Berry curvatures of the respective bands [see Figs.~2(c) and 2(d)]. In fact, the dips and peaks can be enhanced by tuning the Berry curvature with the electric field. The breakings observed after the dips/peaks are due to the spin-splittings of the subbands. They are proportional to the interlayer hopping and strong SOC for holes and they become more distinct for lower temperatures, as shown in Fig.~7(b). For low enough temperatures they evolve into double dips and double peaks, as a consequence of the sharper distribution of the entropy density around the Fermi level. However, in the conduction band the SOC for electrons is much weaker and 
the double-dip (peak) features are not discernible, even at lower temperatures, due to 
the negligible spin-splitting of the subbands. We also notice in Fig.~7(b) the increase in magnitude of the ANC as the temperature is raised. To estimate the Nernst signal at $T=300$ K, 
note that $\alpha_0 = e k_B/4 \pi^2 \hbar\simeq 0.531$ nA/K. 
For the highest peak $\alpha_{xy}^{\text{v}} \simeq 0.5 \alpha_0 = 0.265$ nA/K, which is comparable to $\alpha_{xy}^{\text{v}} $ 
of TMDCs \cite{jauho15,sharma18} and to that of graphene \cite{zhu13}.
\begin{figure}[ptb]
\vspace*{0.2cm}
\begin{center}
\includegraphics[height=3.5cm, width=8.6cm ]{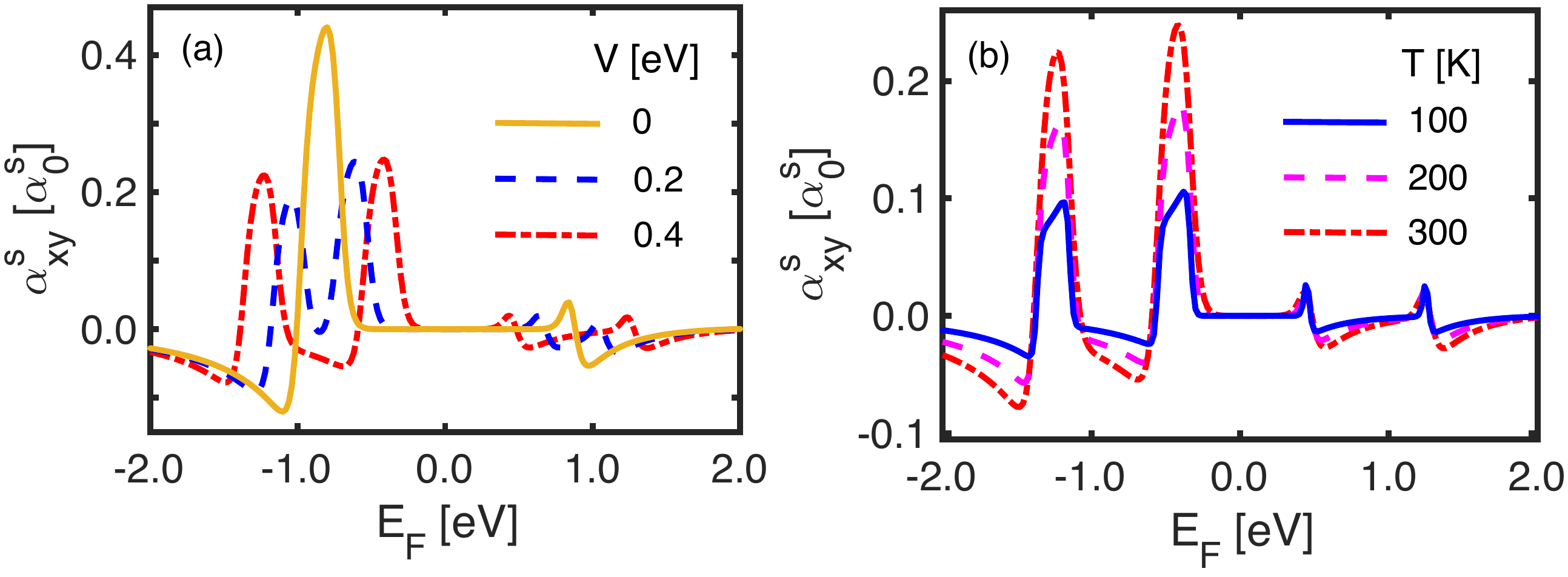}
\end{center}
\vspace*{-0.65cm} \caption{(Colour online) (a) SNC $\alpha_{xy}^{\text{s}}$ for the $K$ valley versus 
$E_F$ for increasing values of electrostatic potential $V$, in units $\alpha_0^{s} = \alpha_0 ( \hbar / 2 e )$. The temperature is $T = 300$ K. (b) The same as in (a) for various temperatures and fixed $V = 0.4$ eV. }
\label{fig:fig1}%
\end{figure} 

In Fig.~8(a) we show the SNC for the $K$ valley versus 
$E_F$ for increasing values of $V$. The temperature is $T = 300$ K. For $V = 0$, the inversion symmetry $P$ is not broken and therefore $S_{\lambda \mu}^{\downarrow} ( \mathbf{k} ) = S_{\lambda \mu}^{\uparrow} ( \mathbf{k} )$. However, $\Omega_{\lambda \mu}^{\uparrow} ( \mathbf{k} ) - \Omega_{\lambda \mu}^{\downarrow} ( \mathbf{k} )$ and hence $\alpha_{xy}^{\text{s}}$, given in Eq.~(\ref{eq296}), is nonzero. This is allowed because both the spin current and in-plane electric field transform in the same manner under $TR$ 
and $P$ 
symmetries \cite{murakami03}. The SNC exhibits  dips and peaks due to the energy shift of the 
valence bands  that have the signs and are proportional to 
the magnitudes of the respective Berry curvatures. 
However, this feature appears inverted in the conduction band, i.e., a peak becomes  a dip, while, at the same time it is significantly degraded due to the absence of interlayer hopping and the much weaker SOC for electrons. For $V > 0$ we observe two dip-peak features in the valence band due to 
the splitting of the spin 
subbands. They 
are suppressed as a result of the smaller Berry curvatures of the spin-down valence bands [see Fig.~2(c)]. The effect of temperature is shown in Fig.~8(b) where we plot the SNC versus $E_F$  
for $V = 0.4$ eV and various temperatures. As expected, the peaks and dips weaken significantly  as the temperature decreases. To estimate the SNC at room temperature 
and $V = 0$, the peak is $\alpha_{xy}^{\text{s}} \simeq 0.45 \times 2 \alpha_0 ( \hbar / 2 e ) = 0.24 ( \hbar / e ) $ nA/K (2 for valley degeneracy).

At low temperatures, Eq.~(\ref{eq295}) can be approximated by the semiclassical Mott relation \cite{cutler69},
\begin{equation}
\alpha_{xy}^{\text{v}} = - \frac{\pi^2 k_B^2 T}{3 e} \frac{d \sigma_{xy}^{\text{v}} ( E_F )}{d E_F}   ,
\label{eq2940}%
\end{equation}
where
\begin{eqnarray}
\nonumber \hspace*{-0.0cm}
\sigma_{xy}^{\text{v}} = \frac{e^2}{\hbar} \sum_{\lambda \mu} \int \frac{d^2 k}{( 2 \pi )^2} \Big[ f_{\lambda \mu}^{\uparrow} ( \mathbf{k} ) \Omega_{\lambda \mu}^{\uparrow} ( \mathbf{k} ) + f_{\lambda \mu}^{\downarrow} ( \mathbf{k} ) \Omega_{\lambda \mu}^{\downarrow} ( \mathbf{k} ) \Big] ,\\*
\label{eq2950}%
\end{eqnarray}
is the valley Hall conductivity. Differentiating Eq.~(\ref{eq2950}) with respect to chemical potential $E_F$, we can rewrite Eq.~(\ref{eq2940}) as
\begin{eqnarray}
\nonumber \hspace*{0in} \alpha_{xy}^{\text{v}} = - \alpha_0 \left( \frac{\pi^2}{3} \right) \sum_{\lambda \mu} \int d^2 k \Big[ f_{\lambda \mu}^{\uparrow} ( \mathbf{k} ) \left( 1 - f_{\lambda \mu}^{\uparrow} ( \mathbf{k} ) \right) \Omega_{\lambda \mu}^{\uparrow} ( \mathbf{k} )  
\\* &&\hspace*{-2.2in} + f_{\lambda \mu}^{\downarrow} ( \mathbf{k} ) \left( 1 - f_{\lambda \mu}^{\downarrow} ( \mathbf{k} ) \right) \Omega_{\lambda \mu}^{\downarrow} ( \mathbf{k} ) \Big] .
\label{eq2000}%
\end{eqnarray}
The Mott relation has been successfully applied in the description of thermoelectric transport in graphene systems \cite{sarma09,tahir15,wei09,zuev09,ma11} even at higher temperatures and in agreement with experimental data \cite{nam10}. In Fig.~9(a) we compare the valley ANC calculated from Eq.~(\ref{eq2000}) with that obtained from the exact equation Eq.~(\ref{eq295}). We use $V = 0.4$ eV and $T = 20$ K. We notice that there is perfect agreement even though the temperature is not very low. The Mott relation predicts that the thermoelectric conductivities have linear temperature dependence for low temperatures. This is shown in Fig.~9(b) where we plot the valley ANC $\alpha_{xy}^{\text{v}}$ versus temperature for $E_F = 0.5$ eV using the Mott relation and compare it with Eq.~(\ref{eq295}). Even though the Mott relation is valid only for low temperatures, we notice that it agrees with Eq.~(\ref{eq295}) up to $\approx170K$. A similar agreement was reported in Ref.~\cite{sarma09} for grapheme monolayers. At higher temperatures, $\sigma_{xy}^{\text{v}} ( E_F )$ becomes smooth, and the contribution of the derivative $d \sigma_{xy}^{\text{v}} / d E_F$ in the Mott relation becomes smaller leading to slower (sublinear) increase of $\alpha_{xy}^{\text{v}}$.
\begin{figure}[ptb]
\vspace*{0.2cm}
\begin{center}
\includegraphics[height=3.5cm, width=8.6cm ]{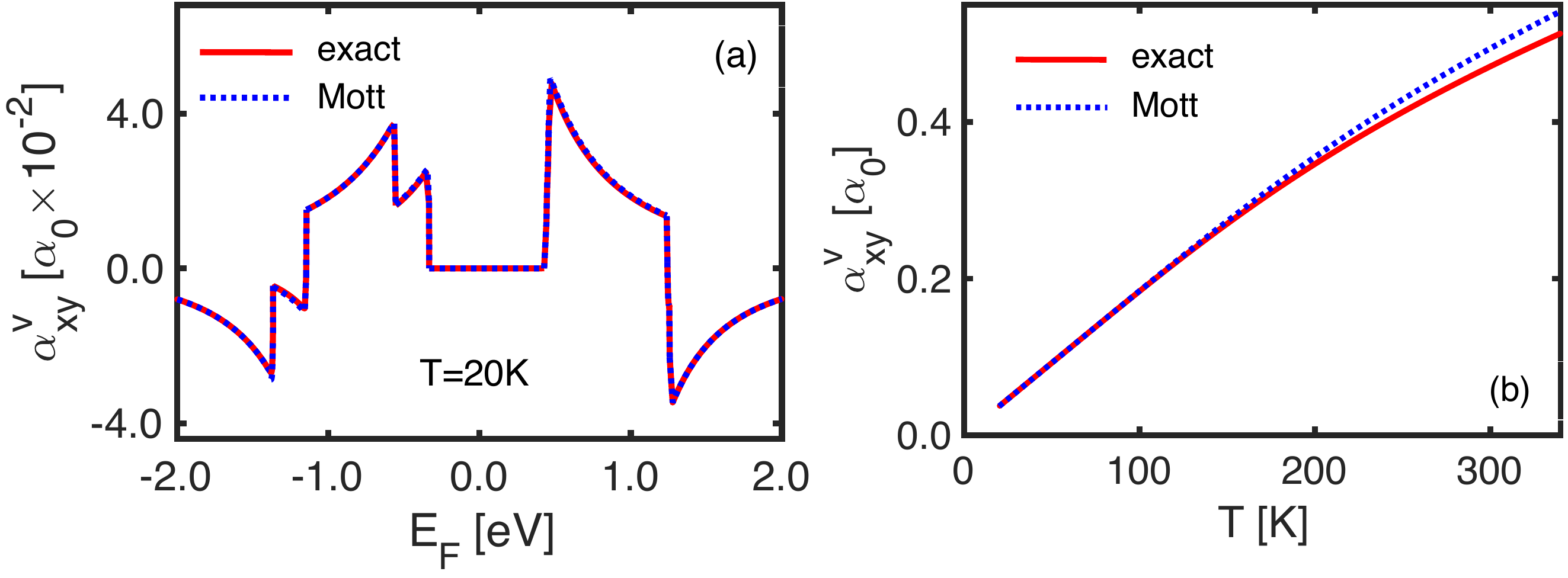}
\end{center}
\vspace*{-0.65cm} \caption{(Colour online) (a) ANC $\alpha_{xy}^{\text{v}}$ for the $K$ valley versus
$E_F$ for $V = 0.4$ eV and $T = 20$ K. Notice the perfect agreement of the result from the exact equation Eq.~(\ref{eq295}) with that from the Mott relation Eq.~(\ref{eq2000}). (b) ANC $\alpha_{xy}^{\text{v}}$ for the $K$ valley versus 
$T$ for $V = 0.4$ eV and $E_F = 0.5$ eV. The deviation from the Mott result starts at $\approx 170$ K. }
\label{fig:fig1}%
\end{figure}

\section{Summary}

Summarizing, we studied the Berry curvature, the OM, 
and the ANE in a biased 
bilayer WSe$_2$. Our results demonstrate that $P$ symmetry breaking by a perpendicular electric field can be used to control the Berry curvature and orbital magnetic moment with important consequences for the OM and the ANE. 
In the absence of an electric field both the OM and the ANC vanish in a particular valley due to the combined effect of $TR$ and $P$ symmetries. In the presence of an electric field they become finite due to the lifting of the spin degeneracy of the bands. 
 In particular, we found that the magnetization is constant and small when the chemical potential $E_F$ is below (above) the lowest (highest) valence (conduction) band. This is because the valence bands of the top and bottom layer carry opposite circulating currents that 
 almost cancel when 
 $E_F$  is in these regions. In-between the bands the magnetization varies linearly with $E_F$ 
 due to a strong imbalance of opposite circulating currents, and is constant in the band gap. The later occurs because
 the derivative of the OM, 
 with respect to $E_F$, 
 is proportional to the Chern number, which is zero for a band insulator.

The 
electric field can generate finite Nernst signals, which exhibit peaks and dips as $E_F$ 
is varied, and thus can be tuned by  it. 
For relatively low temperatures double peaks and double dips 
are clearly observed in the valence band due to the spin splitting and the strong SOC. For zero electric field the SNC 
can be nonzero with its magnitude   larger in the valence band due to the interlayer hopping and strong SOC for holes. 
The  magnitudes of the  dip-peak features are proportional to those of the Berry curvatures of the respective bands. The validity of the semiclassical Mott relation was also verified in a range of temperatures. Our work highlights the role of a gate electric field on certain valley-dependent,  topological  properties and it is pertinent to other bilayer TMDCs. A method to measure the population imbalance at different valleys of TMDCs has been developed recently, thus making possible the realization of the findings of this work \cite{ho20}.

\section*{Acknowledgments}

V.V. and N.N. acknowledge funding from the European Research Council (ERC) under the European Union’s Horizon 2020 Research and Innovation Programme (Grant Agreement No. 678763).

\end{document}